% 2/20/96

\magnification=\magstep 1 
\overfullrule=0pt 
\hfuzz=16pt 
\voffset=0.0 true in 
\vsize=8.8 true in 
\baselineskip 20pt 
\parskip 6pt 
\hoffset=0.1 true in 
\hsize=6.3 true in 
\nopagenumbers 
\pageno=1 
\footline={\hfil -- {\folio} -- \hfil} 
\def\NI{\noindent}
\def\NP{\vfill\eject}

\long\def\UN#1{$\underline{{\vphantom{\hbox{#1}}}\smash{\hbox{#1}}}$}

\ 

\

\centerline{\UN{\bf Exact Solution of an Irreversible One-Dimensional}}

\vskip 0.2in

\centerline{\UN{\bf Model with Fully Biased Spin Exchanges}}

\vskip 0.4in

\centerline{{\bf Ant\'{o}nio M.~R.~Cadilhe}\ \ and\ \ {\bf Vladimir
Privman}}

\vskip 0.2in

\centerline{\sl Department of Physics, Clarkson University, Potsdam,
New York 13699--5820, USA}

\vskip 0.8in

\centerline{\bf ABSTRACT}

\vskip 0.4in

We introduce a model with conserved dynamics, where nearest neighbor
pairs of spins $\uparrow
\downarrow\,(\downarrow \uparrow)$ can exchange to assume the
configuration $\downarrow \uparrow
\,(\uparrow \downarrow)$, with rate $\beta\, (\alpha)$, through energy
decreasing
moves only. We report exact solution for the case when one of the rates,
$\alpha$ or $\beta$, is zero.
The irreversibility of such dynamics results in strong dependence
on the initial conditions.
Domain wall arguments suggest that for more general
models with steady states
the dynamical critical exponent for the anisotropic
spin exchange is different from the isotropic value.

\NP

\NI {\bf 1. INTRODUCTION}

\

Dynamics of one-dimensional systems has received much attention
recently.$^{1-3}$
One-dimensional models can be solved exactly in some cases,
and they provide
convenient test cases for theories of reactions, deposition,
ordering. Furthermore,
in many instances, such as, for instance, diffusion-limited
reactions, low-dimensional 
models offer examples on non-mean-field fluctuations.$^4$
The scope of models studied recently
has been extended significantly$^1$ from the ``classical''
Glauber$^{5}$ and Kawasaki$^{6}$
models of the Ising-spin flips and exchanges.

One of the interesting recent developments has been the introduction of
anisotropic (spatially biased)
dynamical moves in irreversible models of reactions.$^{7-12}$
Anisotropic particle motion also plays key role in hard-core-particle
``asymmetric exclusion'' models.$^{13-14}$ Steady states of kinetic
models with anisotropy have been studied extensively in the field of
driven diffusive systems$^{15}$. Dynamics without detailed balance has
also been considered recently in the field of neural networks.$^{16}$
Our aim in this article has been to initiate investigation of the effect
of making pair exchanges anisotropic in Kawasaki-type spin exchange
models of dynamical behavior.

Most of our results apply in the fully irreversible limit, with
only energy-lowering
moves allowed. This corresponds to the zero-temperature limit
and, in the
case of Kawasaki spin exchanges, yields models of freezing.
Studies of
irreversible low-temperature kinetic Ising models, both spin-flip,
Glauber-,
and spin-exchange, Kawasaki-type, have attracted attention
recently.$^{1}$
The emphasis has been on derivation of exact results for these
and certain
related reaction models. In particular, models with conserved
order parameter, and related systems, were solved by a method
which will be also employed here.$^{17-20}$
Various exactly solvable one-dimensional models of
freezing have been treated by other approaches.$^{21-23}$

Contrary to non-conserved dynamics models of the Glauber type
(spin flip),
where several exact results are available, conserved dynamics,
Kawasaki type models
have long resisted any direct analytical solution. Few exact
results have become available recently for variants in the
zero temperature limit.$^{17-18,24}$
The ordered domain structure in these $T=0$ models becomes
frozen at large times.

Irreversible dynamics of the ``freezing'' type
usually retains infinite memory of the initial
conditions. We will use two different
initial conditions, namely, the fully alternating
lattice corresponding to a concentration of $1\over 2$ of both
spin types, and
random initial conditions where each site is either occupied by
a $\uparrow$ spin
with probability $p$ or by a $\downarrow$ spin with probability $q=1-p$.

The outline of the paper is as follows: We start by
defining the model in
Section~2. Then in Section~3 we present the exact solution for the
single rate (fully anisotropic) case. Finally, in Section~4 we discuss
the results and
consider generally the anisotropic exchange models in the framework
of domain-wall arguments
which, for the $T>0$ dynamics, yield information on the dynamical
critical exponent.$^{21,25-27}$

\NP

\NI {\bf 2. THE MODEL}

\

When two spins are exchanged in a ferromagnetic-interaction model,
the total
energy can be increased, unchanged, or decreased.
By ferromagnetic interaction
we mean that the energy of a ``bond'' connecting parallel spins
($\uparrow
\uparrow$ or $\downarrow\downarrow$) is lower than the energy of a bond
connecting antiparallel spins ($\uparrow\downarrow$
or $\downarrow\uparrow$).

In order to simplify the notation, we will use the language of
particle exchanges from now on. Thus, the
$\uparrow$ ($\downarrow$) spins will be
replaced by
particles $A$ ($B$). We consider only the dynamics
in the $T\to 0$ limit and allow
for energy-decreasing particle exchanges. Such dynamics leads
to frozen states.$^{17-18,20}$

Consider a one-dimensional lattice where each site is occupied
by particle $A$ or $B$.
The only two possible nearest-neighbor particle exchanges that lower the
energy locally are
$$\cdots ABAB \cdots \, \to
\cdots AABB \cdots \qquad ({\rm rate\ } \alpha) \eqno(1)$$
$$\cdots BABA \cdots \, \to
\cdots BBAA \cdots \qquad ({\rm rate\ } \beta) \eqno(2)$$
Note that the particle (spin) exchanges are anisotropic. The usual
isotropic Kawasaki-model spin exchanges correspond to the equal rates
$\alpha = \beta$. For finite temperatures, exchanges other than those
shown in (1)-(2) are allowed, and their relative rates depend on the
temperature.$^{6}$ In the Ising model notation, exchanges (1)-(2)
lower the energy by $4J$, where $-J$ is the energy of the $AA$ and $BB$
bonds, while $J>0$ is the energy of the $AB$ and $BA$ bonds.

The model defined by the moves (1)-(2) is conveniently analyzed$^{18}$
in terms of the probabilities $P_{C,n} (t)$ that a randomly chosen
interval of $n>1$ lattice sites is fully alternating, $ABAB\cdots$ or
$BABA\cdots$, at time $t$. Here $C \in \{A,B\}$ denotes the starting
site type in the $n$-site sequence. For $n>3$, the interior of such
intervals is ``fully reactive,'' i.e., all internal
nearest-neighbor exchanges are allowed (reduce energy).
In order to be able to claim that the probabilities $P_{C,n} (t)$ are
independent of the position along the one-dimensional lattice,
we must assume that
the initial conditions are uniform; specific choices will be
specified later (see also Section 1).
It is important to emphasize that the
probabilities
only refer to the interior of the interval which might or might not be 
a part of a larger alternatively-ordered interval. The probabilities are
\UN{\sl not\/} conditioned on the arrangement of particles outside the
selected $n$-site sequence. 

A closed hierarchy of rate equations can now be written as
follows. Consider first the case of even $n=2k$, where $k>1$.
The rate equation for $P_{A,2k}(t)$ is obtained in the form
$$ -\dot P_{A,2k}=(k-1)\alpha P_{A,2k} + (k-2) \beta P_{A,2k}
+\beta P_{A,2k+1} + \beta P_{B,2k+1} + 2 \alpha P_{A,2k+2} \eqno(3) $$
Here the dot denotes the time derivative. The first term in (3)
represents the rate at which the $k-1$ sequences of the type (1),
internal to the selected interval of $2k$ consecutive sites,
undergo the process given in (1) thereby destroying the fully
alternatively-ordered state in the selected interval. Similarly,
the second term corresponds to the $k-2$ different four-site
subsequences in the selected interval, which undergo process (2).

The third and the fourth terms correspond to, respectively,
the rightmost and the leftmost $AB$ pairs in the selected
interval of $2k$ sites, exchanging as parts of four-site sequences
of type (2). In such a process the four-site sequence
containing the end-pair of the selected interval must in fact
be a part of a larger, ($2k+1$)-site, ordered sequence. Thus,
$P_{C,2k+1}$ is used. Finally, the last term corresponds
to processes in which the two end sites of the selected interval
exchange with sites external to it. The configuration is of
type (1) at both ends, and the alternating order must extend to
a larger interval of $2k+2$ sites, in order for process (1) to
occur. 

Similarly, we get
$$ -\dot P_{B,2k}=(k-1)\beta P_{B,2k} + (k-2) \alpha P_{B,2k}
+\alpha P_{B,2k+1} + \alpha P_{A,2k+1} + 2 \beta P_{B,2k+2} \eqno(4) $$
It is also of interest to write down the rate equation for the case
$n=2$. Indeed, the quantities $P_{C,2}$ yield densities of $AB$
and $BA$ interfaces in the system. The interfaces separate ordered
domains of $A$ and $B$ particles ($\uparrow$ and $\downarrow$
spins). We have
$$ -\dot P_{A,2}=2\alpha P_{A,4} + \beta P_{B,4} \eqno(5) $$
$$ -\dot P_{B,2}=2\beta P_{B,4} + \alpha P_{A,4} \eqno(6) $$

We next turn to the case of odd $n=2k+1$,
with $k \geq 1$. The rate equations now read
$$ -\dot P_{A,2k+1}=(k-1)\alpha P_{A,2k+1} + (k-1) \beta P_{A,2k+1}
+\beta P_{B,2k+2} + \alpha P_{A,2k+2} + (
\alpha + \beta ) P_{A,2k+3} \eqno(7) $$
$$ -\dot P_{B,2k+1}=(k-1)\beta P_{B,2k+1} + (k-1) \alpha P_{B,2k+1}
+\alpha P_{A,2k+2} + \beta P_{B,2k+2} + (
\alpha + \beta ) P_{B,2k+3} \eqno(8) $$

We must complement the above system of rate equations with initial
conditions.
For the initially alternating lattice of $A$ and $B$ particles,
the probability of starting at either site type is $1/2$ regardless
of the interval size, i.e.,
$$ P_{A,2k}^{\rm alt}(0)=P_{B,2k}^{\rm alt}(0)=
P_{A,2k+1}^{\rm alt}(0)=P_{B,2k+1}^{\rm alt}(0)=
{1 \over 2} \eqno(9) $$
For the initially random distribution, placing $A$ particles with
probability $p$ and $B$ particles with probability $q=1-p$, at each
lattice site, the initial conditions are
$$ P_{A,2k}^{\rm ran}(0)=P_{B,2k}^{\rm ran}(0)=(pq)^k \eqno(10) $$
$$ P_{A,2k+1}^{\rm ran}(0)=p(pq)^k \eqno(11) $$
$$ P_{B,2k+1}^{\rm ran}(0)=q(pq)^k \eqno(12) $$

The above system of rate equations turns
out to be exactly solvable in some
cases. The case $\alpha=\beta$ (isotropic exchanges) has
been considered in the literature.$^{17-18,24}$ 
We could not find a solution
for the general case of both rates being
non-zero and not equal to each other. In the next section we solve the
case of $\alpha>0$ and $\beta=0$. We note, however, that the
transformation
used to eliminate the $k$-dependence (see the next section) can be
generalized to the case of  $\alpha>0$, $\beta>0$. Thus, a system of
ordinary differential equations can be obtained and analyzed numerically
or by approximation methods.

\NP

\NI {\bf 3. THE SINGLE RATE CASE}

\   

In this section we consider a solvable case of the fully anisotropic 
exchanges. Specifically, we take $\beta=0$, $\alpha>0$, so that only
exchanges of the type (1) are allowed. It is furthermore convenient to
redefine the time variable (or alternatively,
set $\alpha=1$) in such a way that
$$ t_{\rm new} = \alpha t_{\rm old} \eqno(13) $$
This eliminates the $\alpha$-dependence from the rate equations.

More importantly, the $k$-dependence can be also eliminated,
by a transformation
which generalizes the one used in the isotropic case.$^{17-18,24}$ 
Thus, we try a solution of the form 
$$ P_{A,2k} (t)=S(t) \gamma^k(t) \eqno(14) $$
$$ P_{B,2k} (t)=U(t) \gamma^k(t) \eqno(15) $$
$$ P_{A,2k+1} (t)=V(t) \gamma^k(t) \eqno(16) $$
$$ P_{B,2k+1} (t)=W(t) \gamma^k(t) \eqno(17) $$
Direct substitution in (3)-(4), (7)-(8) yields
(with $\alpha=1$, $\beta=0$)
$$ \dot \gamma=-\gamma \eqno(18) $$
$$ \dot S=(1-2\gamma) S \eqno(19) $$ 
$$ \dot U=2U-(V+W) \eqno(20) $$     
$$ \dot V=(1-\gamma) V-\gamma S \eqno(21) $$
$$ \dot W=(1-\gamma) W-\gamma S \eqno(22) $$
We must also check that the proposed
forms (14)-(17) are consistent with the
initial conditions. Denoting the initial values by the subscript 0, we
identify
$$ \gamma_0^{\rm alt}=1 \eqno(23) $$
$$ S_0^{\rm alt}=U_0^{\rm alt}=V_0^{\rm alt}=W_0^{\rm alt}
={1\over 2} \eqno(24) $$
and
$$ \gamma_0^{\rm ran}=pq \eqno(25) $$
$$ S_0^{\rm ran}=U_0^{\rm ran}=1 \eqno(26) $$
$$ V_0^{\rm ran}=p \eqno(27) $$
$$ W_0^{\rm ran}=q \eqno(28) $$
These two types of initial conditions were introduced in Section~2.

For both initial conditions the solution can be now
obtained straightforwardly. First, we solve (18),
$$ \gamma(t)=\gamma_0 e^{-t} \eqno(29) $$
This result is then used to solve (19),
$$ S(t)=S_0 \exp\left[2\gamma_0 e^{-t}-2\gamma_0+t\right] \eqno(30) $$
The next step is to solve the equations for $V$ and $W$, (21)-(22),
$$ V(t)=(V_0-S_0)\exp\left[\gamma_0
e^{-t}-\gamma_0+t\right]+S_0\exp\left[2\gamma_0
e^{-t}-2\gamma_0+t\right] \eqno(31) $$
$$ W(t)=(W_0-S_0)\exp\left[\gamma_0
e^{-t}-\gamma_0+t\right]+S_0\exp\left[2\gamma_0
e^{-t}-2\gamma_0+t\right] \eqno(32) $$
Finally, the equation for $U$ is solved,
$$ U(t)= \left(U_0-{V_0+W_0-S_0 \over \gamma_0}\right) e^{2t}
+{V_0+W_0-2S_0 \over \gamma_0}
\exp\left[\gamma_0 e^{-t}-\gamma_0+2t\right]$$
$$+{S_0 \over \gamma_0}\exp\left[2\gamma_0 e^{-t}-2\gamma_0+2t\right]
\eqno(33) $$

These expressions provide the full, time-dependent solution
of the problem.
Specifically,
the probabilities $P_{C,n} (t)$ are obtained from (30)-(33),
with (14)-(17),
for all $n \geq 3$. However, for $n=2$ special expressions
apply,
$$P_{A,2} (t)=P_{A,2} (0)+{S_0
\gamma_0}\exp\left[2\gamma_0 e^{-t}-2\gamma_0\right]
-S_0 \gamma_0  \eqno(34) $$
$$P_{B,2} (t)=P_{B,2}
(0)+{S_0 \gamma_0\over 2}\exp\left[2\gamma_0 e^{-t}-2\gamma_0\right]
-{S_0 \gamma_0\over 2}  \eqno(35) $$
These results were obtained by
integrating (5)-(6), with $\alpha=1$, $\beta=0$.

\NP

\NI {\bf 4. DISCUSSION}

\

As in the isotropic-exchange case,$^{17-18,24}$ the most profound 
feature of the dynamics
is its strong dependence on the initial conditions.
In fact, the initial values are ``remembered''
for all times, and enter the $t \to \infty$ limiting
expressions. The large-time limiting
expressions can be obtained by expanding in powers
of the quantity $\gamma_0 e^{-t}$ 
$\, \left( \, \equiv \gamma (t) \, \right)$ which
enters in the exponentials in (31)-(35). 
In fact, for large times we have $S,V,W \propto e^{t}$,
$U \propto e^{2t}$, where the proportionality
constants depend on the initial values. Thus, only the shortest
alternatively-ordered intervals
survive at large times. Specifically,
$P_{A,2} (\infty)$, $P_{B,2} (\infty)$,
$P_{A,3} (\infty)$, $P_{B,3} (\infty)$, $P_{B,4} (\infty)$ ``survive'' 
(remain nonzero) for the fully anisotropic
dynamics. All other probabilities vanish exponentially.

The final configuration is therefore incompletely ordered.
Some interfaces remain at $t=\infty$.
These are pairs $AB$ and $BA$, the probability of which
is given by $P_{A,2}$ and $P_{B,2}$,
respectively. Some of the interfaces survive as nearby
pairs, represented by the nonzero values
of $P_{A,3}$ and $P_{B,3}$. Furthermore, some triplets of interfaces
remain unreacted, due to $\beta=0$, corresponding
to configuration (2). The probability of the latter
is given by $P_{B,4}$.
For illustration, let us calculate the total density
of interfaces per site, $I(t)$, in the system, which
was also obtained in the isotropic-model studies.$^{17-18,24}$ We get 
$$ I(t)=P_{A,2} (t)+P_{B,2} (t) \eqno(36)$$
$$ I^{\rm alt}(t)={1 \over 4 } \left\{ 1 + 3 \exp \left[ 2 \left( e^{-t}
- 1 \right) \right] \right\} \eqno(37)$$
$$ I^{\rm ran}(t)={pq \over 2 } \left\{ 1 + 3 \exp \left[ 2pq
\left( e^{-t}
- 1 \right) \right] \right\} \eqno(38)$$
We note that both the large-time values themselves, and
the coefficients of the exponential
approach to these values, depend on the initial conditions.

For the initially fully alternating lattice, it is interesting to
notice that 
$$ P_{A,2}^{\rm alt}(t) = P_{A,3}^{\rm alt}(t)= P_{B,3}^{\rm alt}(t) =
 P_{B,4}^{\rm alt}(t)
={1 \over 2 }\exp \left[ 2 \left( e^{-t}- 1 \right) \right]
\eqno(39)$$
For this particular case, the lattice can be viewed as initially fully
covered by $AB$ ``objects.'' Each \UN{\sl nearest-neighbor pair\/}
of such objects can ``react away'' via the process (1).
At the end of the process, all the \UN{\sl original\/} unreacted $AB$
pairs remain in the centers of configurations $BABA$. This process is
therefore identical to Random Sequential Adsorption of dimers on the
linear lattice,$^{28}$ and in fact one can derive the expression for
$P_{B,4}(t)$ from the exact expression available for the adsorption of
dimers. However, other quantities in (39) are less straightforward; we
were not able to propose a ``dynamical'' argument for the equalities
(39).

We turn next to domain wall arguments as applied to more
general one-dimensional models
with anisotropic spin exchanges. We will not specify the model but
only assume that the $\uparrow$ spins favor to move to the left, while
the $\downarrow$ spins are favorably moved to the right. In the
particle notation, $BA \to AB$ exchanges are favored over $AB \to BA$.

For very low
but finite temperatures, the configuration mainly consists of single
isolated interfaces, separating large ordered domains, at the distance
of the correlation length, $\xi$. If we
denote by $\Delta E=2J >0$ the energy cost to create an interface
by ``breaking''
a ferromagnetic bond, then for \UN{\sl equilibrium states\/} entropy
arguments$^{29}$ suggest that
$$ \xi \sim \exp \left[ \Delta E /k_BT \right] 
=  \exp \left[ 2J /k_BT \right] \eqno(40) $$
for low temperatures. Of course, with anisotropic spin exchanges,
the system may not have an equilibrium state but rather reach a steady
state$^{15}$ at large times. The identification (40) then
must be checked by exact or approximate methods.

The usual domain wall argument$^{21,25-27}$ for isotropic
conserved order parameter models at low temperatures, can
be phrased as follows. The slowest relaxation mode in the system will
be a locally correlated, diffusional motion of
isolated interfaces, on the time
scale proportional to the inverse of the
rate of diffusion and to the separation
squared, i.e., to $\xi^2$. However, the rate of diffusion is 
rather small. An isolated interface cannot move on its own by spin
exchanges. It must first be turned into a triplet
of interfaces by a spin exchange of energy cost $4J$. Thus, the
diffusion rate will be proportional to the inverse Boltzmann factor,
$\exp \left[ -4J/k_BT \right]$.

In addition, a pair of
interfaces will ``unbind'' and diffuse away at the rate of order 1
(because spin exchanges that lead to motion of a pair of interfaces
do not cost energy). This pair must reach a neighboring isolated
interface
at a distance of order $\xi$ for the process to be complete. If it
instead returns to the interface from which it originated, then the
configuration will be restored to the original one. The probability of
not returning back is $\sim
1/\xi ; {}^{30}$ this yields another factor of
$\xi$ in the relaxation time.$^{21,25}$
The resulting expression for the relaxation time $\tau$ is written as
$$ \tau_{\rm isotropic} \sim \xi^5 \eqno(41) $$
where we used (40). This yields the value $z=5$ for the dynamical
critical exponent in $\tau \propto \xi^z$.

For anisotropic exchange, the unbinding interface pairs will be
immediately returned back for ``attracting'' interfaces, but they will
move away by biased diffusion (with asymptotically negligible
probability of return) for ``repelling'' interfaces. (The two types of
interfaces alternate in
the system.) Therefore, the probability factor of
not returning back, the inverse of $1/\xi$,
will not be present. We expect
$$ \tau_{\rm anisotropic} \sim \xi^2 \exp \left[ 4J/k_BT \right]
\eqno(42) $$
If (40) applies for the specific model's steady state, then we identify
$z=4$, a different value as compared to the isotropic case.

In summary, we reported new exact results for the anisotropic exchange
model with only the most energy-favorable moves allowed. This
restriction allows exact solvability in the fully anisotropic case. The
resulting dynamics is that of freezing. Domain wall arguments
suggest that for more realistic, $T>0$ models with similar dynamical
rules, there will be new interesting effects such as the new value of
the dynamical critical exponent.

\NP

\NI {\bf REFERENCES}

\ 

{\frenchspacing

\item{1.} V. Privman, editor, {\sl Nonequilibrium Statistical
Mechanics in One Dimension\/}
(Cambridge University Press, Cambridge, 1996), in print.

\item{2.} J. Stat. Phys. {\bf 65}, 
nos. 5/6 (1991), Proceedings of {\sl Models of Non-Classical
Reaction Rates}, NIH (March 25-27, 1991).

\item{3.} V. Privman, {\sl Dynamics of Nonequilibrium
Processes: Surface Adsorption, Reaction-Diffusion Kinetics, Ordering
and Phase Separation}, in {\sl Trends in Statistical
Physics}, Vol. 1, p. 89  (Council for Scientific Information,
Trivandrum, India, 1994).

\item{4.} Review: S. Redner, Ch. 1 in Ref. 1.

\item{5.} R. Glauber, J. Math. Phys. {\bf 4}, 294 (1963).

\item{6.} K. Kawasaki, Phys. Rev. {\bf 145}, 224 (1966).

\item{7.} V. Privman, J. Stat. Phys. {\bf 72}, 845 (1993).

\item{8.} S. A. Janowsky, Phys. Rev. E{\bf 51}, 1858 (1995).
 
\item{9.} S. A. Janowsky, {\sl Spatial Organization in the Reaction
$A+B\to$ inert for Particles with a Drift}, preprint.
 
\item{10.} I. Ispolatov, P. L. Krapivsky, and S. Redner,
{\sl Kinetics of $A+B\to 0$ with Driven Diffusive Motion}, preprint.
 
\item{11.} V. Privman, E. Burgos, and M. Grynberg, 
Phys. Rev. E{\bf 52}, 1866 (1995).
 
\item{12.} V. Privman, A. M. R. Cadilhe, and M. L. Glasser,
J. Stat. Phys. {\bf 81}, 881 (1996).
 
\item{13.} Review: S.A. Janowsky and J.L. Lebowitz, Ch. 13 in Ref. 1.

\item{14.} Review: B. Derrida and M.R. Evans, Ch. 14 in Ref. 1.

\item{15.} B. Schmittmann and R.K.P. Zia, in
{\sl Phase Transitions and Critical Phenomena}, edited by C. Domb and
J.L. Lebowitz, Vol. 17 (Academic, London, 1995).

\item{16.} N. Mousseau and D. Sherrington, J. Phys. A{\bf 28},
6557 (1995).

\item{17.} Review: V. Privman, Mod. Phys. Lett. B{\bf 8}, 143 (1994).

\item{18.} V. Privman, Phys. Rev. Lett. {\bf 69}, 3686 (1992).

\item{19.} S.N. Majumdar and V. Privman, J. Phys. A{\bf 26}, L743
(1993).

\item{20.} S. N. Majumdar and C. Sire, Phys. Rev. Lett. {\bf 70},
4022 (1993).

\item{21.} Review: S.J. Cornell, Ch. 6 in Ref. 1.

\item{22.} S.J. Cornell, K. Kaski and R.B. Stinchcombe, Phys. Rev.
B{\bf 44}, 12263 (1991).

\item{23.} R. Schilling, J. Stat. Phys. {\bf 53}, 1227 (1988).

\item{24.} J.-C. Lin and P.L. Taylor, Phys. Rev. E{\bf 48}, 4305
(1993).

\item{25.} R. Cordery, S. Sarker and J. Tobochnik,
Phys. Rev. B{\bf 24}, 5402 (1981).

\item{26.} W. Zwerger, Phys. Lett. {\bf 84}A, 269 (1981).

\item{27.} U. Decker and F. Haake, Z. Physik B{\bf 35}, 281 (1979).

\item{28.} Review: J.W. Evans, Ch. 10 in Ref. 1.

\item{29.} V. Privman and M.E. Fisher, J. Stat. Phys. {\bf 33}, 385
(1983).

\item{30.} M.E. Fisher, IBM J. Res. Dev. {\bf 32}, 76 (1988).

}
 
\bye